\documentclass[conference]{IEEEtran}
\IEEEoverridecommandlockouts

\usepackage{cite}
\usepackage{amsmath,amssymb,amsfonts}
\usepackage{algorithmic}
\usepackage{graphicx}
\usepackage{textcomp}
\usepackage{xcolor}
\usepackage{booktabs,hyperref}
\def\BibTeX{{\rm B\kern-.05em{\sc i\kern-.025em b}\kern-.08em
    T\kern-.1667em\lower.7ex\hbox{E}\kern-.125emX}}
\newcommand{\etal}{\textit{et al}.}
\begin{document}

\title{Advancing NAM-to-Speech Conversion with Novel Methods and the MultiNAM Dataset}

\author{\IEEEauthorblockN{Neil Shah\IEEEauthorrefmark{1}\IEEEauthorrefmark{2},
Shirish Karande\IEEEauthorrefmark{2} and Vineet Gandhi\IEEEauthorrefmark{1}}

\IEEEauthorblockA{\IEEEauthorrefmark{1} CVIT, IIIT Hyderabad, India \mbox{ } \mbox{ } \IEEEauthorrefmark{2} TCS Research Pune, India\\
Email: neilkumar.shah@research.iiit.ac.in, shirish.karande@tcs.com, vgandhi@iiit.ac.in}}
\maketitle

\IEEEpubid{
    \begin{minipage}{\textwidth}
    \scriptsize
    \raggedright
    Copyright 2025 IEEE. Published in ICASSP 2025 – 2025 IEEE International Conference on Acoustics, Speech and Signal Processing (ICASSP), scheduled for 6-11 April 2025 in Hyderabad, India. Personal use of this material is permitted. However, permission to reprint/republish this material for advertising or promotional purposes or for creating new collective works for resale or redistribution to servers or lists, or to reuse any copyrighted component of this work in other works, must be obtained from the IEEE. Contact: Manager, Copyrights and Permissions / IEEE Service Center / 445 Hoes Lane / P.O. Box 1331 / Piscataway, NJ 08855-1331, USA. Telephone: + Intl. 908-562-3966.
    \end{minipage}
}

\IEEEpubidadjcol

\begin{abstract}
Current Non-Audible Murmur (NAM)-to-speech techniques rely on voice cloning to simulate ground-truth speech from paired whispers. However, the simulated speech often lacks intelligibility and fails to generalize well across different speakers. To address this issue, we focus on learning phoneme-level alignments from paired whispers and text and employ a Text-to-Speech (TTS) system to simulate the ground-truth. To reduce dependence on whispers, we learn phoneme alignments directly from NAMs, though the quality is constrained by the available training data. To further mitigate reliance on NAM/whisper data for ground-truth simulation, we propose incorporating the lip modality to infer speech and introduce a novel diffusion-based method that leverages recent advancements in lip-to-speech technology. Additionally, we release the MultiNAM dataset with over $7.96$ hours of paired NAM, whisper, video, and text data from two speakers and benchmark all methods on this dataset. Speech samples and the dataset are available at \url{https://diff-nam.github.io/DiffNAM/}
\end{abstract}


\section{Introduction}
\label{sec:intro}

\IEEEpubidadjcol

Silent Speech Interfaces (SSIs) are devices that enable communication without audible speech~\cite{DENBY2010270}. SSIs capture physiological signals, such as muscle movements, associated with speech production and convert them into speech. SSIs are crucial for individuals who have lost their speaking abilities, such as those who have undergone a laryngectomy \cite{Janke} and are also useful in acoustically harsh environments or where silence is necessary, like hospitals or quiet public spaces.

SSI techniques capture articulator movements using various sensors and imaging techniques such as Ultrasound tongue imaging~\cite{HUEBER2010288}, Electromyography~\cite{SCHULTZ2010341}, real-time MRI~\cite{otani23_interspeech}, Electrolarynx~\cite{espy1998enhancement}. However, these techniques are constrained by their invasiveness~\cite{SCHULTZ2010341} or the need for highly sensitive tracking devices~\cite{HUEBER2010288}. Nakajima \etal~\cite{nakajima2003non} introduced a method to capture NAMs (signals lacking acoustic intelligibility and incomprehensible even to nearby listeners) from tissues behind the ear using a specialized microphone. This SSI technique offers advantages like content privacy, availability in select markets, good performance in noisy environments, and cost effectiveness~\cite{DENBY2010270}. Subsequent efforts improved device design and usability~\cite{nakajima2006methods} and studied the sensitivity and frequency characteristics of several NAM microphones~\cite{shimizu2007acoustic, shimizu2009frequency}. Yang \etal~\cite{yang2012noise} also released a $40$-minute corpus from a single speaker of paired NAM, whispered speech, and corresponding text to aid further research.

\IEEEpubidadjcol

Given NAM vibrations and the corresponding ground-truth speech, the task can be framed as a direct sequence-to-sequence translation. However, the unavailability of ground truth speech remains the primary hurdle. By definition, since the subjects only murmur, at most, only the whisper sound can be captured. A simple approach to obtain ground-truth speech is to record clean speech in a studio and then use Dynamic Time Warping to align it with input NAMs~\cite{malaviya2020mspec}. However, warping techniques introduce unnatural distortions in the aligned signals, resulting in poor intelligibility of the converted speech. 
Shah~\etal~\cite{shah24_interspeech} leverage self-supervised methods to simulate ground-truth speech by converting whispers into speech. While promising, this approach has its limitations—specifically, the simulation method does not generalize well to whisper data from different speakers. Furthermore, the reliance on paired whisper data presents a significant hurdle, as such data may not always be accessible, especially from patients with speech difficulties. To address this dependency, one alternative involves deriving phoneme-level alignments between NAMs and text and feeding these durations into a TTS module~\cite{stethospeech}. However, within a resource-scarce setting where NAM data is limited, the resulting alignments tend to be noisy. Additionally, current methods and datasets overlook the potential of visual modalities, which could offer a fresh perspective.

\IEEEpubidadjcol

To address the aforementioned limitations, we present a MultiNAM dataset, containing $7.96$ hours of NAM vibrations with corresponding text, whispers, and facial videos. We benchmark existing methods alongside newly proposed methodologies in various scenarios where different input modalities are available, such as whisper and text, NAMs and text, whisper alone, or only facial video, among others. We study these methods in both high-resource and resource-scarce settings. To eliminate reliance on paired whispers and improve intelligibility in resource-scarce scenarios, we explore existing state-of-the-art (SOTA) lip-to-speech synthesis methods and introduce a novel diffusion model conditioned on simulated NAMs and video to generate ground-truth speech. 

\section{Datasets}
\begin{figure}[t]
  \centering
  \includegraphics[width=0.96\linewidth]{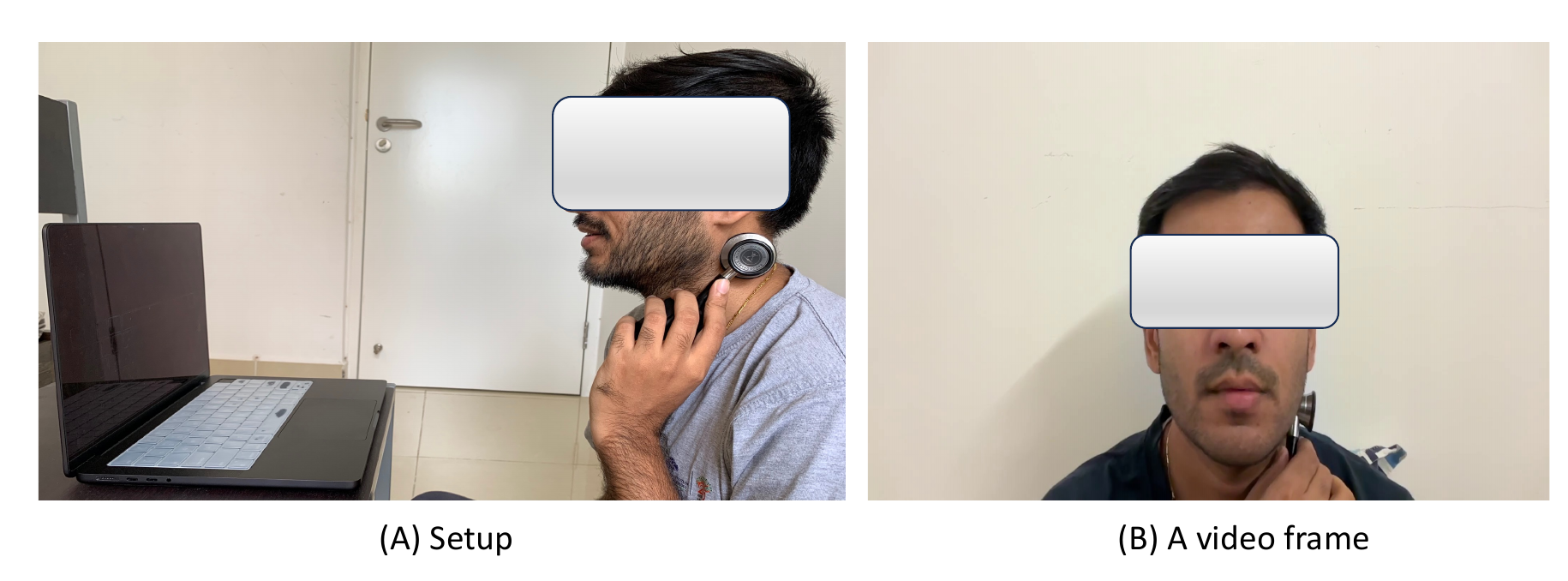}
  \caption{Data recording setup: A laptop displays text while recording the speaker’s face and whispering voice. A stethoscope head placed behind the ear captures NAM vibrations.}
  \label{fig:setup}
\end{figure}

The proposed MultiNAM corpus was collected in a typical office-like environment. Fig.~\ref{fig:setup} (A) demonstrates the recording setup. In an effort towards making the technology more accessible, we employed an off-the-shelf Bluetooth-connected digital stethoscope~\footnote{https://www.ayusynk.ai} for recording the NAM vibrations, instead of using specially fabricated NAM microphones, as in prior art. We record the NAM vibrations, the murmuring sound (whisper), and the face video with the corresponding text. The subjects were asked to position the stethoscope head on the tissue area behind their ears, following recommendations from previous studies~\cite{nakajima2003non,nakajima2006methods,stethospeech}. We collected data from two subjects, aged 31—one male ($S1$) and one female ($S2$). They were instructed to read sentences, borrowed from the LJSpeech~\cite{ljspeech17} dataset, in a soft, whisper voice, with each participant assigned different text to ensure distinct contributions. Speaker $S1$ provided $2,354$ utterances, while speaker $S2$ narrated $969$ sentences, resulting in a total of $5.66$ hours and $2.30$ hours of paired NAM, facial video, whisper, and text data, respectively. Both participants had no known hearing or speech impairments and exhibited no signs of pathology. Video feeds focusing on their faces and corresponding whisper recordings from the mouth area were captured using a MacBook Pro notebook, at a resolution of $1920$ x $1080$ pixels and $25$ fps. The whispers and NAMs were recorded at a sampling frequency of $48$ kHz. 

To examine the frameworks in a resource-limited scenario and their sensitivity to high-quality whisper speech, we additionally experiment with the CSTR NAM TIMIT Plus corpus \cite{yang2012noise}. This dataset comprises $421$ studio-recorded sentences from a female speaker, including NAM vibrations paired with whisper audio and text, spanning $40$ minutes at a $16$ kHz sampling frequency. We use OpenAI's Whisper speech recognizer~\cite{pmlr-v202-radford23a} to calculate error rates. The whisper speech in the CSTR NAM TIMIT Plus corpus has a Word Error Rate (WER) of $1.07\%$, while our corpus shows $5.54\%$ for speaker $S1$ and $12.97\%$ for speaker $S2$. Raw NAMs have a WER of $66.99\%$ in the CSTR corpus and $200.50\%$ and $192.77\%$ for speakers $S1$ and $S2$, respectively, in our corpus.

\section{Method}
Our method follows a two-step approach (a) simulating aligned ground-truth speech corresponding to NAMs and (b) learning a Sequence-to-Sequence (Seq2Seq) model to convert NAMs into speech. In this section, we primarily focus on various methodologies for simulating aligned ground-truth speech. We use an approach similar to~\cite{shah24_interspeech} for Seq2Seq modeling given the simulated speech. 

\subsection{Ground-truth speech simulation from whisper}
\label{sec: is24}
Shah \etal~\cite{shah24_interspeech} relied on HuBERT~\cite{hsu2021hubert} as a speech encoder to obtain content-rich representations that focus on ignoring speaker and ambient information, rather than using Mel-cepstral features for encoding whisper. During training, a HiFi-GAN~\cite{kong2020hifi} speech decoder is trained on HuBERT representations derived from the LJSpeech dataset. During inference, HuBERT representations from the available whisper are processed through the trained speech decoder to generate an aligned ground-truth speech. Since whispers and NAMs are aligned, this process creates a paired corpus of input NAMs and simulated ground-truth speech using the available paired whisper. We refer to this method as \textit{HuBERT-HiFi}.

\subsection{Ground-truth simulation using forced alignment}
\label{sec: mfa}
Shah \etal~\cite{stethospeech} proposed \textit{StethoSpeech}, an alternative approach to generating aligned ground-truth speech by learning phoneme-level alignments between the input NAM and the target text to estimate the duration of each phoneme. These durations are then explicitly fed into a TTS model aligned with the text to simulate ground-truth speech. In this work, we experiment with both NAMs and whispers in correspondence with the target text to synthesize ground-truth speech and validate the efficacy of the method. To achieve this, we train an acoustic model on the provided audio-text pairs to determine phoneme durations. We train the acoustic model using the Montreal Forced Aligner (MFA)~\cite{mcauliffe2017montreal} and employ FastSpeech2~\cite{ren2021fastspeech} as the TTS model. We refer to this method as \textit{MFA-TTS}.

\subsection{Ground-truth simulation Using vision modality}
We address the problem of simulating ground-truth speech as a lip-to-speech synthesis task, generating speech from silent video. In this section, we explore two recent methods: \textit{Cross-Attention}~\cite{Sindhuliptospeech} and \textit{Lipvoicer}~\cite{yemini2024lipvoicer}. Since these methods require ground-truth speech for training, we cannot train them directly on our data. Therefore, we train the models using the LRS3~\cite{afouras2018lrs3} dataset and then apply these trained models to our videos during inference.

{\bf Data Preprocessing:} Following most TTS works \cite{parrottts,ren2021fastspeech}, we convert text into phonemes. For every video segment, we locate the $68$ facial landmarks using dlib~\cite{dlib} and align each frame to a reference face frame with an affine transformation as described in~\cite{shi2022learning}. We then crop an $88$×$88$ lip region centered on the mouth and convert each frame to grayscale. To extract lip representations, we employ AV-HuBERT~\cite{shi2022learning} as our video encoder. Using all the available (video, text) pairs from our dataset, we fine-tune the entire AV-HuBERT model for visual speech recognition with Connectionist Temporal Classification (CTC)~\cite{graves2006connectionist} loss and extract lip features from the final layer of the fine-tuned model.

\noindent {\bf Cross-Attention:} We follow the architecture proposed by Sindhu~\etal~\cite{Sindhuliptospeech} but differ in preprocessing. The system takes text and video from the LRS3 dataset as input, with speech as the training target. The phonemes are processed by a text encoder, while the AV-HuBERT lip features are encoded into visual embeddings using a visual encoder. Both encoders are based on transformer layers from~\cite{vaswanitransformer}. To achieve video-text temporal alignment, we use multi-head scaled-dot product attention~\cite{vaswanitransformer}, where visual embeddings serve as queries and text embeddings act as both keys and values. We use a pre-trained HuBERT speech encoder to obtain the target speech representation. This choice is motivated by the similarity in training methodologies between AV-HuBERT and HuBERT, which we believe will yield content-rich representations conducive to improved intelligibility. The HuBERT model encodes speech at $50$ Hz and AV-HuBERT at $25$ Hz, indicating a temporal relationship. Thus, target HuBERT durations can be obtained by upscaling AV-HuBERT durations by $2$ before passing through a transformer decoder. The transformer decoder predicts HuBERT speech units, optimized with cross-entropy loss. These units are then passed through a pre-trained HiFi-GAN speech decoder to generate speech.

\noindent {\bf Diffusion Process:} \textit{Lipvoicer}, a recent lip-to-speech method~\cite{yemini2024lipvoicer} trains a conditional denoising diffusion probabilistic model (DDPM)~\cite{kong2021diffwave} on video to generate mel-spectrograms. The generation process during inference is guided by the video and either predicted transcriptions from a pre-trained lip reading network or, if available, ground-truth transcriptions. However, the speech lacks intelligibility and quality in our videos, whether using predicted text or available ground-truth text. We propose modifying the training approach to generate mel-spectrograms by conditioning the DDPM on both video and synthetic NAMs. Given a training data point sampled from the data distribution $x_0 \sim p_{\text{data}}(x)$ and $\beta_t$ a pre-defined noise schedule, DDPM defines a forward process $q(x_t | x_{t-1}) = \mathcal{N}\left(x_t; \sqrt{(1 - \beta_t)}x_{t-1}, \beta_t I\right)$ that iteratively turns input into Gaussion noise and a reverse process~\cite{NEURIPS2020_4c5bcfec} is then learned by a network that approximates $q(x_{t-1}|x_t,x_0)$. 

Many diffusion models use classifier guidance for conditional generation, but recent approaches~\cite{yemini2024lipvoicer} improve performance with classifier-free guidance~\cite{ho2021classifierfree}, which we also employ in training our diffusion model. We extract $16$ kHz speech signals from the training videos of the LRS3 dataset and use these to generate mel-spectrograms. During training, the DDPM generates $1$-second mel-spectrograms conditioned on the video and synthetic NAM vibrations, using $L_{1}$ loss for diffusion noise prediction. To generate synthetic NAMs for speech samples from the LRS3 dataset, we train a HiFiGAN speech decoder using our recorded NAMs. After training, we input the HuBERT representations of LRS3 speech samples into the trained speech decoder to produce NAMs that reflect the content of the LRS3 speech. Unlike~\cite{yemini2024lipvoicer}, which relies on pre-trained ShuffleNet v2, Temporal Convolutional Network, and ResNet-18 for video feature extraction, we use fine-tuned AV-HuBERT representations for lip video and NAMs to emphasize content. We avoid conditioning on image embeddings to prevent the model from learning speaker-specific information. While retaining the same DDPM backbone as~\cite{yemini2024lipvoicer}, we adjust several parameters: the hop length is set to $320$, and the window length to $800$. Lip video embeddings are concatenated with NAM embeddings, resulting in a final conditioner size of $1536$. All other architectural parameters remain unchanged from~\cite{yemini2024lipvoicer}. During inference, we condition the model on available ground-truth text instead of lip-reading transcriptions~\cite{yemini2024lipvoicer} to effectively handle unconstrained vocabulary. We refer to this method as \textit{Diff-NAM}.

\subsection{Seq2Seq learning network and speech decoder}
\label{sec: seq2seq}
Utilizing ground-truth speech using the approaches above, we train a non-autoregressive transformer~\cite{ren2021fastspeech} to model the relationship between the latent spaces of NAMs and the ground truth. The $6$-layer Seq2Seq network employs feed-forward transformer blocks with two multi-head self-attention layers and two $1D$ convolutions, similar to FastSpeech2~\cite{ren2021fastspeech}. The encoder converts NAM vibrations into fixed-dimensional vectors, while the decoder generates ground-truth speech embeddings. The model optimizes the mean squared error loss, quantifying the difference between the decoded and ground-truth speech embeddings. Following~\cite{shah24_interspeech}, we add a fully connected layer after the transformer encoder to enhance text prediction by optimizing CTC loss. We train a modified HiFiGAN-v2~\cite{polyak21_interspeech} with a batch size of $16$, a learning rate of $2 \times 10^{-4}$, $100$ embeddings, an embedding dimension of $128$, and an input dimension of $256$. The HiFiGAN speech decoder synthesizes speech from embeddings predicted by the Seq2Seq network.

\begin{table}[t]
\caption{Recognition performance on simulated ground-truth speech using whisper modality across different approaches}
\begin{center}
    \begin{tabular}{cccccc}
    \toprule
    Dataset & Speaker & \multicolumn{4}{c}{Method}                                    \\
            \cmidrule(lr){1-6}
            &         & \multicolumn{2}{c}{\textit{HuBERT-Hifi}~\cite{shah24_interspeech}} & \multicolumn{2}{c}{\textit{MFA-TTS}~\cite{stethospeech}} \\
            &         & WER             & CER           & WER          & CER          \\
            \cmidrule(lr){3-4} \cmidrule(lr){5-6}
    CSTR    & -       & \textbf{23.77}           & \textbf{14.00}         & 41.07        & 24.22        \\
    \midrule
    Ours    & $S1$      & 100.14          & 63.83         & \textbf{5.84}         & \textbf{2.71}         \\
    Ours    & $S2$      & 84.53           & 51.76         & \textbf{11.23}        & \textbf{4.93}      \\
    \bottomrule
    \end{tabular}
\label{tab:whisper}
\end{center}
\end{table}

\begin{table*}[t]
    \caption{Error rates of simulated ground-truth and converted speech using lip/nam modalities across different approaches}
\begin{center}
    \begin{tabular}{ccclll|llll}
    \toprule
    Method                 & Modality             & \multicolumn{4}{c}{Simulated speech}                     & \multicolumn{4}{c}{Converted speech}                                                                  \\
    \midrule
                           &                      & \multicolumn{2}{c}{$S1$}          & \multicolumn{2}{c|}{$S2$} & \multicolumn{2}{c}{$S1$}                            & \multicolumn{2}{c}{$S2$}                            \\
    \multicolumn{1}{l}{}   & \multicolumn{1}{l}{} & \multicolumn{1}{l}{WER} & CER   & WER        & CER       & \multicolumn{1}{c}{WER} & \multicolumn{1}{c}{CER} & \multicolumn{1}{c}{WER} & \multicolumn{1}{c}{CER} \\
    \cline{3-10}
    \textit{MFA-TTS (all speaker)}~\cite{stethospeech} & NAM, text            & \textbf{12.37}                   & \textbf{9.16}  & \textbf{19.64}      & \textbf{14.19}     & \textbf{26.38}                   & \textbf{15.12}                   & \textbf{32.61}                   & \textbf{18.97}                   \\
    \textit{MFA-TTS (per-speaker)}~\cite{stethospeech}               & NAM, text            & 23.81                   & 13.15 & 33.62      & 19.13     & 56.65                   & 35.74                   & 69.45                   & 48.31                   \\
    \textit{Cross-Attention}~\cite{Sindhuliptospeech}        & Lip, text            & 67.54                   & 39.43 & 62.13      & 35.64     & 93.04                   & 59.45                   & 87.98                   & 61.39                   \\
    \textit{Lipvoicer (GT)}~\cite{yemini2024lipvoicer}         & Lip, text            & 27.01                   & 18.37 & 33.94      & 21.99     & 54.11                   & 31.16                   & 59.66                   & 35.10                   \\
    \textit{Lipvoicer (Pred)}~\cite{yemini2024lipvoicer}       & Lip, text            & 39.04                   & 27.02 & 49.99      & 33.13     & 74.51                   & 58.32                   & 79.46                   & 62.38                   \\
    \textbf{\textit{Diff-NAM (Ours)}}               & Lip, NAM, text       & \textbf{17.24}                   & \textbf{11.31} & \textbf{21.73}      & \textbf{13.97}     & \textbf{32.39}                   & \textbf{19.47}                   & \textbf{38.94}                   & \textbf{24.91}      \\
    Mspec-Net \cite{malaviya2020mspec}              &          -             &             -             &             -             &      -     &     -      &
    141.23                  & 69.62                   & 156.19                  & 98.23      \\
    \bottomrule
    \end{tabular}
\label{tab: nowhisper}
\end{center}
\end{table*}

\section{Results and Discussion}
\subsection{Ground-truth speech recognition with whisper}
\label{sec: withwhisper}
We compare the performance of the simulated ground-truth speech from \textit{HuBERT-Hifi}~\cite{shah24_interspeech} and \textit{MFA-TTS}~\cite{stethospeech} methods trained with paired whisper data. The results are presented in Table~\ref{tab:whisper}. The \textit{HuBERT-Hifi} method achieves high intelligibility (WER: $23.77\%$) when trained on whisper data from the CSTR corpus, but its performance degrades with our whisper data (WER: $100.14\%$ for speaker $S1$). This suggests poor generalization to diverse speakers, likely due to the HuBERT model's limited exposure to regional accents. The \textit{MFA-TTS} method shows improved performance for speaker $S1$ compared to speaker $S2$ and those in the CSTR corpus, indicating that it requires more whisper-text pairs for training; performance drops when less data is available.

\subsection{Ground-truth speech recognition without whisper}
\label{sec: nowhisper}
\textit{MFA-TTS} performs exceptionally well when a large amount of whisper data is available (WER: $5.84\%$ for speaker $S1$). However, in many practical scenarios (e.g., patients with voice disorders), whisper data is not always accessible. Therefore, in this section, we explore experiments where ground-truth speech was simulated without using whisper data. The results are presented in Table~\ref{tab: nowhisper}. The ``Modality" column indicates the input modality used for training or inference. We observed that in this setting, \textit{MFA-TTS (all speaker)} method, trained on NAM vibrations and text from both speakers, achieves the best performance. If the alignment is done on individual speakers (\textit{MFA-TTS (per-speaker)}), we observe a drop in performance, with WER increasing from $12.37\%$ to $23.81\%$ for speaker $S1$. This strengthens our earlier observation (see Section~\ref{sec: withwhisper}) that increased data enhances forced alignment, resulting in more intelligible ground-truth speech.

The techniques explored (\textit{MFA-TTS} and \textit{HuBERT-Hifi}) require the speaker's audio data (NAMs/whispers) for training. However, in extreme cases, such as patients with chronic obstructive pulmonary disease \cite{barnes2003chronic}, where airflow is severely restricted and the muscles are weakened to the point of being unable to murmur or whisper, these approaches may fail to effectively simulate ground truth. Therefore, we investigate techniques for simulating ground-truth speech without relying on the audio modality and experiment with lip modality alone by using lip-to-speech models trained on large out-of-domain datasets. We used the \textit{Cross-Attention}~\cite{Sindhuliptospeech} method trained on the LRS3 dataset, but it resulted in the highest error rates, indicating its ineffectiveness for our in-the-wild videos. We then apply the pre-trained diffusion-based lip-to-speech model \cite{yemini2024lipvoicer} to infer ground-truth speech from our recorded videos. When using lip-reader predicted text as a conditioning factor alongside video during inference, error rates ranged from $39.04\%$ to $49.99\%$ for speaker $S1$ and $S2$. In contrast, conditioning with ground-truth text—available during our ground-truth simulation—reduced error rates to $27.01\%$ for speaker $S1$ and $33.94\%$ for speaker $S2$. Our proposed Diff-NAM method further enhanced performance, lowering error rates to $17.24\%$ for $S1$ and $21.73\%$ for $S2$.  This improvement underscores the key role of content-specific preprocessing and DDPM conditioning with simulated NAMs in generating more accurate ground-truth speech.

\noindent {\bf NAM-to-speech conversion without whisper:} Table~\ref{tab: nowhisper} also shows the error rates of converted speech from mapping input NAMs via Seq2Seq network. The term ``converted speech" refers to the final output of the NAM-to-Speech process. These results validate our hypothesis that the quality and intelligibility of the ground truth impact the intelligibility of the converted speech. The converted speech using the simulated ground-truth from the \textit{MFA-TTS} baseline performs well with ample NAM data but struggles in resource-scarce scenarios, with WER increasing from $26.38\%$ to $56.65\%$ for speaker $S1$. Speech converted using simulated ground-truth from lip-to-speech methods (\textit{Cross-Attention} and \textit{Lipvoicer}) trained on out-of-domain datasets results in poor intelligibility, with the WER remaining above $50\%$. In contrast, speech converted using the ground truth simulated by our proposed \textit{Diff-NAM} method achieves the lowest error rates ($32.39\%$ for $S1$ and $38.94\%$ for $S2$) among all lip-to-speech methods. This demonstrates that improving task-specific preprocessing and conditioning the diffusion process with simulated NAMs yields significantly better results.

\section{Conclusion}
This study proposes to eliminate the reliance on paired whisper data for ground-truth speech simulation. Existing \textit{HuBERT-Hifi} method struggle to generalize to new speakers even when using paired whisper data. The \textit{MFA-TTS} baseline based on the StethoSpeech method achieves the lowest error rates with available paired whispers. Using \textit{HuBERT-Hifi} and \textit{MFA-TTS} with ample NAMs and corresponding text improves performance and eliminates the need for paired whisper data. However, these methods, when trained with limited samples, struggle with intelligibility. To eliminate reliance on NAMs/whispers for ground-truth simulation, we leverage advances in lip-to-speech synthesis and propose a novel diffusion model. We introduce a $7.96$-hour multi-modal dataset with paired NAMs, whispers, facial videos, and texts from two speakers to support further research. Our future goal is to refine cross-modality alignments, crucial for boosting the intelligibility of simulated ground-truth speech and advancing Seq2Seq network training.

\bibliographystyle{IEEEtran}
\bibliography{refs}
\end{document}